\documentclass{egpubl}
\usepackage{egsr18}

\usepackage{booktabs} 

\usepackage{import}
\usepackage[super,negative]{nth}

\usepackage[ruled]{algorithm2e}

\ifpdf \usepackage[pdftex]{graphicx} \pdfcompresslevel=9

\else \usepackage[dvips]{graphicx} \fi

\PrintedOrElectronic

\usepackage{t1enc,dfadobe}

\usepackage{egweblnk}
\usepackage{cite}

\def\fg#1{Fig.~\ref{fig:#1}}

\def\boldsymbol#1{\textbf{#1}}

\def\CS{StyleBlit}

\title[\CS]%
      {\CS: Fast Example-Based Stylization with Local Guidance}

\iffalse
\author[paper1034]{\parbox{\textwidth}{\centering paper1034}}
\else
\author[S\'{y}kora et al.]
{\parbox{\textwidth}{\centering D. S\'{y}kora$^{1,2}$, O. Jamri\v{s}ka$^{1}$, J. Lu$^{3}$, E. Shechtman$^{3}$ }\\
{\parbox{\textwidth}{\centering
         $^1$Czech Technical University in Prague, Faculty of Electrical Engineering, Czech Republic\\
         $^2$University of Utah, USA\\
         $^3$Adobe Research, USA
         }
}
}
\fi

\begin{document}

\teaser{\def\svgwidth{\hsize}
\begingroup%
  \makeatletter%
  \providecommand\color[2][]{%
    \errmessage{(Inkscape) Color is used for the text in Inkscape, but the package 'color.sty' is not loaded}%
    \renewcommand\color[2][]{}%
  }%
  \providecommand\transparent[1]{%
    \errmessage{(Inkscape) Transparency is used (non-zero) for the text in Inkscape, but the package 'transparent.sty' is not loaded}%
    \renewcommand\transparent[1]{}%
  }%
  \providecommand\rotatebox[2]{#2}%
  \ifx\svgwidth\undefined%
    \setlength{\unitlength}{1860.64765625bp}%
    \ifx\svgscale\undefined%
      \relax%
    \else%
      \setlength{\unitlength}{\unitlength * \real{\svgscale}}%
    \fi%
  \else%
    \setlength{\unitlength}{\svgwidth}%
  \fi%
  \global\let\svgwidth\undefined%
  \global\let\svgscale\undefined%
  \makeatother%
  \begin{picture}(1,0.24037156)%
    \put(0,0){\includegraphics[width=\unitlength]{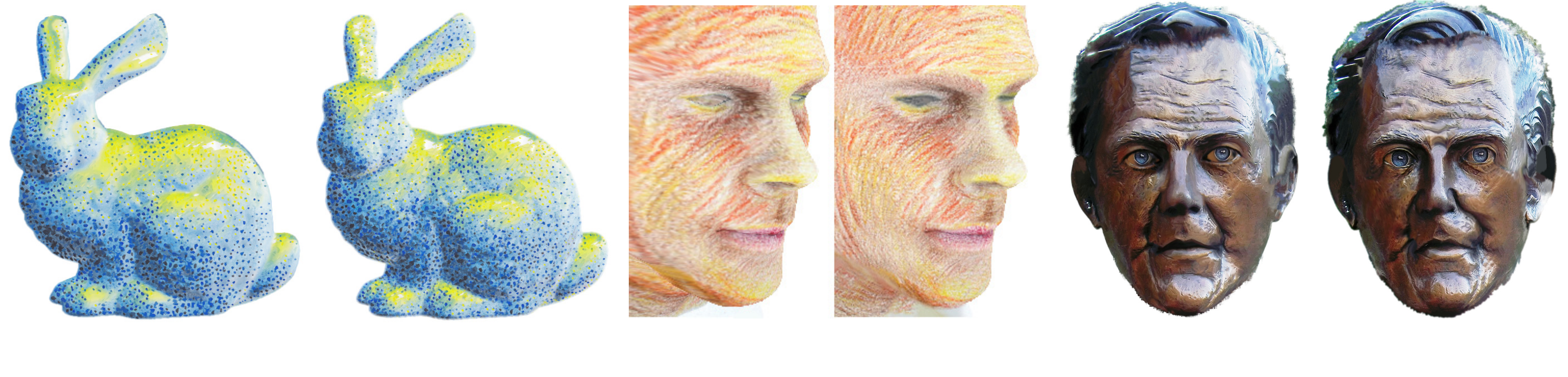}}%
    \small
    \put(0.09952470,0.015){\color[rgb]{0,0,0}\makebox(0,0)[b]{\smash{(a)~StyLit [\boldsymbol{56 secs}]}}}%
    \put(0.29572581,0.015){\color[rgb]{0,0,0}\makebox(0,0)[b]{\smash{(b)~our approach [\boldsymbol{0.05 sec}]}}}%
    \put(0.59580168,0.015){\color[rgb]{0,0,0}\makebox(0,0)[b]{\smash{(d)~our approach}}}%
    \put(0.46569363,0.015){\color[rgb]{0,0,0}\makebox(0,0)[b]{\smash{(c)~texture mapping}}}%
    \put(0.75443302,0.015){\color[rgb]{0,0,0}\makebox(0,0)[b]{\smash{(e)~FaceStyle [\boldsymbol{83 secs}]}}}%
    \put(0.91871995,0.015){\color[rgb]{0,0,0}\makebox(0,0)[b]{\smash{(f)~our approach [\boldsymbol{0.1 sec}]}}}%
  \end{picture}%
\endgroup%
\caption{\CS~in applications: (a)~style transfer
from an exemplar in~\fg{res} to a 3D model using StyLit~\protect\cite{Fiser16};
(b)~our approach delivers similar visual quality but is several orders of
magnitude faster; (c)~regular texture mapping using texture presented
in~\fg{tex} vs. (d)~our approach that better preserves visual characteristics
of the used artistic media; (e)~style transfer to a portrait image using
FaceStyle~\protect\cite{Fiser17} with an exemplar in their supplementary
material; (f)~our approach produces similar visual quality and is notably
faster.}\label{fig:teaser}}

\maketitle
\begin{abstract}

We present~\CS---an efficient example-based style transfer algorithm that can
deliver high-quality stylized renderings in real-time on a single-core CPU. Our
technique is especially suitable for style transfer applications that use local
guidance - descriptive guiding channels containing large spatial variations.
Local guidance encourages transfer of content from the source exemplar to
the target image in a semantically meaningful way. Typical local guidance
includes, e.g., normal values, texture coordinates or a displacement field.
Contrary to previous style transfer techniques, our approach does not involve
any computationally expensive optimization. We demonstrate that when local
guidance is used, optimization-based techniques converge to solutions that can
be well approximated by simple pixel-level operations. Inspired by this
observation, we designed an algorithm that produces results visually similar to,
if not better than, the state-of-the-art, and is several orders of magnitude
faster. Our approach is suitable for scenarios with low computational budget
such as games and mobile applications.


\end{abstract}

\section{Introduction}
\label{sec:intro}
\begin{figure*}[ht!]
\def\svgwidth{\hsize}\import{}{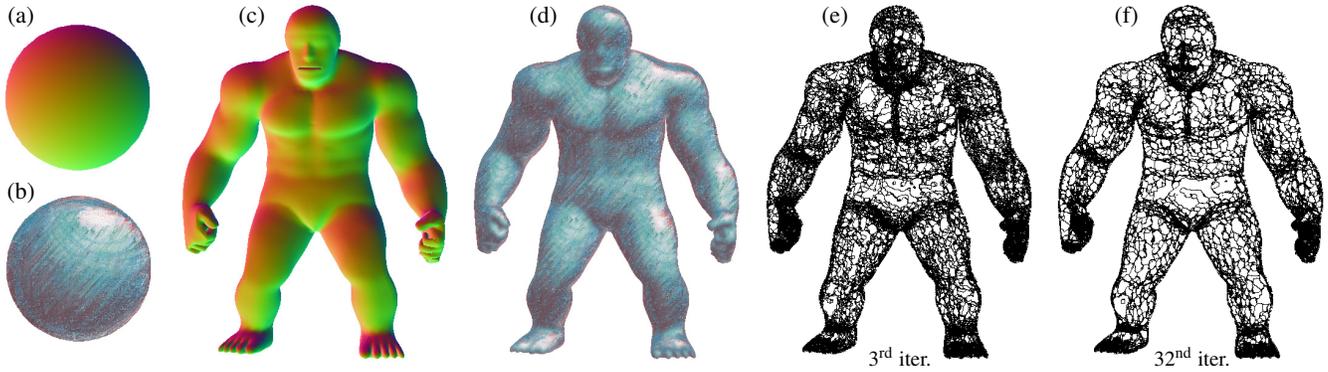}\caption{Motivation for our approach: state-of-the-art
guided patch-based synthesis~\protect\cite{Fiser16} is used to transfer
artistic style from a hand-drawn sphere~(b) onto a more complex 3D object~(c).
Normal maps are used as guidance~(a, c). The result~(d) nicely preserves
textural coherence of the original artistic style since the optimization-based
approach converges to a state where large coherent chunks of the source texture
are copied into the target image forming a mosaic~(e, f).}\label{fig:inspire}
\end{figure*}

Example-based artistic style transfer recently became popular thanks to
advances made by neural-based approaches~\cite{Gatys16,Selim16}, patch-based
texture synthesis techniques~\cite{Fiser16,Fiser17} and their
combinations~\cite{Li16,Liao17}. These methods can produce impressive
stylization results with a common limitation of high computational overhead.
Although interactive frame-rate can be achieved when compromising visual
quality~\cite{Johnson16} or utilizing the GPU~\cite{Fiser16}, high quality
style transfer remains out of reach for scenarios such as interactive games or
mobile applications where the avaialbe computational budget is low.

A key concept that distinguishes example-based style transfer from regular
texture synthesis~\cite{Efros99} is the use of guiding
channels~\cite{Hertzmann01}. Those encourage the transfer of a specific area in
the source exemplar to a corresponding area in the target image. The design of
guiding channels is extremely important for achieving semantically meaningful
transfer. The guidance can be relatively \emph{fuzzy} with respect to a certain
spatial location (e.g., segmentation or blurred gray-scale gradients used by
Hertzmann et al.) or well-localized and descriptive (e.g., a displacement
field~\cite{Selim16,Fiser17}, texture coordinates~\cite{Rematas14,Magnenat15}
or normal values~\cite{Sloan01,Diamanti15}). We call the later \emph{local}
guidance.

The goal of current state-of-the-art patch-based style transfer
algorithms~\cite{Fiser16,Zhou17} is to optimize for a solution that satisfies
the prescribed guidance and consists of large coherent chunks of the style
exemplar in semantically meaningful regions. This ideal solution represents the
most visually-pleasing configuration that maximizes sharpness and fidelity of
the synthesized texture since large areas of the exemplar are copied as is
(see~\fg{inspire}). To achieve this, however, \emph{textural
coherence}~\cite{Kwatra05,Wexler07} needs to be taken into account which
results in a computationally demanding energy minimization problem.

In this paper, we demonstrate that when guiding channels provide good
localization and when style exemplars contains mostly stochastic textures,
textural coherence becomes less important as the local characteristics of the
guide implicitly encourage coherent solutions and the stochastic nature
provides visual masking that suppresses visible seams. In this setting we
demonstrate that expensive optimization can be effectively replaced by a set of
simple and fast pixel-level operations that gain significant performance
speed-up. On a single core modern CPU we can stylize a one-megapixel image at
10 frames per second while on a common GPU we can achieve more than 100 frames
per second at a 4K UHD resolution. Despite its simplicity, our new method
produces high-quality transfer results for a wide range of styles. Applications
include stylization of 3D renderings~\cite{Fiser16} (see~\fg{teaser}, left),
image-based texture mapping that better preserves the characteristics of
natural artistic media~\cite{Magnenat15} (\fg{teaser}, middle), or fast style
transfer to faces with comparable results to the method of Fi\v{s}er et
al.~\shortcite{Fiser17} (\fg{teaser}, right).

\section{Related Work}
\label{sec:related}
\begin{figure*}[ht!]
\def\svgwidth{\hsize}\import{}{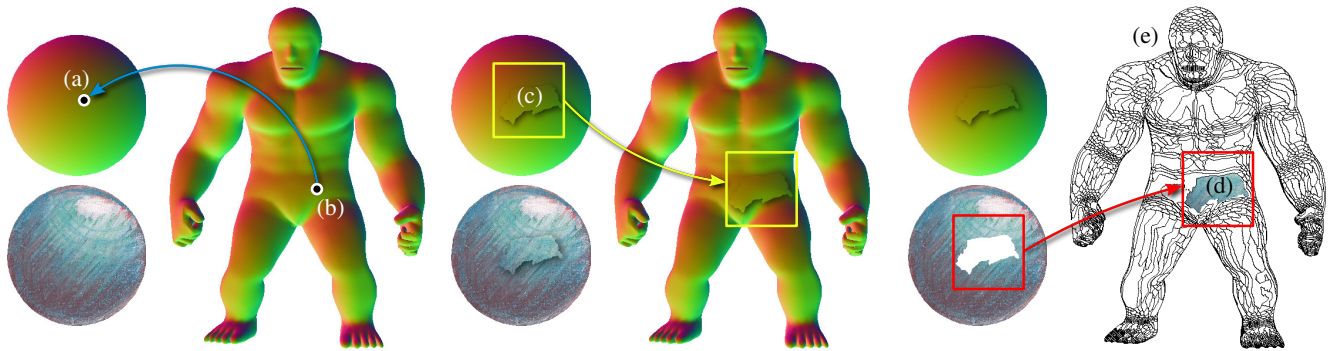}\caption{The core idea behind our method: for each
randomly selected seed in the target image~(b), we perform a table lookup using
its guidance value (in this case a normal) to retrieve the corresponding
location in the source exemplar~(a). Then we compare the guidance values of
source and target pixels in spatially-aligned regions around the seed. Pixels
with a guidance value difference below a user-defined threshold belong to the
same chunk~(c). Finally we transfer the chunk of example pixels to the
target~(d). We can produce the final mosaic by repeating this process~(e).}\label{fig:splat}
\end{figure*}

Over the last two decades, non-photorealistic rendering~\cite{Kyprianidis13}
evolved considerably. The state-of-the-art techniques can synthesize images
resembling real artwork. A popular branch of techniques achieves this goal by
mixing a set of predefined strokes or patterns that are selected and positioned
according to guiding information provided in 2D~\cite{Hertzmann98} or
3D~\cite{Schmid11} environments. In addition to painterly styles, this line of
approaches can also simulate other artistic styles such as pen-and-ink
illustration~\cite{Salisbury97} or hatching~\cite{Breslav07}. Nevertheless,
these approaches are confined by the limited expressive power of these
predefined sets of strokes or patterns.

To alleviate this drawback, an example-based approach called \emph{Image
Analogies} was introduced by Hertzmann et al.~\shortcite{Hertzmann01}. This
method allows an artist to prepare an arbitrary stylized version of a target
image given an input style example. A one-to-one mapping between the input
image and its stylized version is used to guide the transfer by establishing
correspondences between the source and target (based, e.g., on color
correspondence). The target image can then be stylized according to this
analogy. This seminal concept was later extended to animations~\cite{Benard13}
and improved by others~\cite{Barnes17} using better synthesis
algorithms~\cite{Kaspar15,Fiser16} as well as different types of
guidance~\cite{Zhou17,Fiser17}. In parallel, an approach similar to \emph{Image
Analogies} was introduced by Sloan et al.~\shortcite{Sloan01} and later
extended by others~\cite{Barla06,Todo13}. Their technique called~\emph{The Lit
Sphere} (a.k.a.~\emph{MatCap}) uses a one-to-one correspondence between normal
values to transfer style from a hand-drawn exemplar of a simple object (a
sphere) to a more complex 3D model. In this scenario, simple environment
mapping can be used~\cite{Blinn76} to perform the transfer. Recently, Magnenat
et al.~\shortcite{Magnenat15} proposed a similar technique where instead of
normals, UV coordinates are used as guidance so that the artist can draw a
stylized version on a 2D projection of a 3D model and then the style is
transferred using texture mapping. This approach is similar to image-based
texture mapping used in 3D reconstruction~\cite{Debevec96}. Style transfer can
be performed in real-time thanks to its simplicity, but it only works well when
the style does not contain distinct high-frequency details. Texture mapping
often distorts high-frequency details failing to retain the fidelity of the
used artistic medium. It was demonstrated later~\cite{Fiser16,Bi17} that much
better results can be obtained using patch-based synthesis that not only takes
into account local guidance but also textural coherence, however, at the cost
of notably higher computational overhead.

Recently, Gatys et al.~\shortcite{Gatys16} introduced an alternative approach
to style transfer based on parametric texture synthesis~\cite{Portilla00} where
instead of a steerable pyramid, an alternative parametric representation is
used based on a deep neural network trained for object
recognition~\cite{Simonyan14}. Their technique inspired a lot of follow-up
work~\cite{Semmo17} and became very popular thanks to numerous publicly
available implementations. Although it produces impressive results for some
style exemplars, it was shown to suffer from certain high-frequency artifacts
caused by the parametric nature of the synthesis algorithm~\cite{Fiser16,Fiser17}.
To prevent texture distortion, researchers have proposed techniques to combine
the advantages of patch-based synthesis and the deep features learned by neural
network~\cite{Li16,Liao17}. These approaches, however, have large computational
overhead and are not suitable for real-time applications.

Our approach to style transfer bears resemblance to early texture synthesis
approaches~\cite{Praun00,Liang01,Efros01,Kwatra03} that can achieve results
similar to patch-based synthesis~\cite{Kwatra05,Wexler07} by transferring
larger irregularly-shaped chunks of the source exemplar and composing them
seamless in the target image. In particular \emph{Lapped
Textures}~\cite{Praun00} can tile the target surface with a set of source
patches, however, there is no specific guidance for the patch placement, the
patches need to be prepared in advance to have minimal features on boundaries
(to avoid seams), and the approach requires additional growing operation to
fill in gaps. In appearance-space texture synthesis~\cite{Lefebvre06} small
appearance vectors are used instead of color patches to compress neighborhood
information, nevertheless, still an iterative optimization~\cite{Lefebvre05} is
necessary to obtain the final result.

In another related work~\shortcite{Pritch09} a graph labelling problem is
solved to find optimal shift of every pixel in the output image from its source
in an input image. Nevertheless, additional smoothness term is needed to avoid
discontinuities and so computationally demanding optimization is required.

In this paper we demonstrate that for hand-drawn exemplars which are usually
highly stochastic the interplay between local guidance and textural masking
effect described by Ashikhmin et al.~\shortcite{Ashikhmin01} makes seams
between the individual chunks barely visible and thus simple blending operation
can be used to suppress them percetually without the need to take into account
texture coherence explicitly.

\section{Our Approach}
\label{sec:method}

In this section, we describe the core idea behind our approach and discuss
implementation details. As a motivation, we first describe a simple experiment
that inspired us to develop our method.

To understand the properties of optimization-based approaches, we applied the
StyLit algorithm~\cite{Fiser16} to transfer the style from a hand-drawn image
of a sphere to a more complex 3D model using normals as guidance
(see~\fg{inspire}). The texture coherence term in the original energy
formulation, and the mechanism for preventing excessive utilization of source
patches, help the optimization converge to a state where large chunks of the
original source texture (\fg{inspire}b) are copied to the target image
resulting in a high-fidelity transfer (\fg{inspire}d).

Inside each coherent chunks, the errors of texture coherence term are equal to
zero. Errors of the guidance term can be bounded by a small upper bound, i.e.,
we can find a chunk of the normal field on the exemplar sphere to roughly
approximate the corresponding chunk of normals on the target 3D model within a
certain error threshold. The black lines in \fg{inspire}e, f show the
boundaries between chunks within which all pixels have guidance errors below
some predefined error bound. The lines get sparser and the regions grow larger
as the bound increases.

This fact inspired us to seek large coherent chunks of style regions directly
using simple pixel-level operations foregoing expensive patch-based optimization.

\subsection{Basic Algorithm}

To build such a mosaic of coherent chunks, we need to estimate the shape and
spatial location of each individual chunk. This is done by going in a scan-line
order or by picking a random pixel (seed) in the target image and finding its
corresponding location in the source exemplar (see~\fg{splat}a, b). Usually the
local guidance at each target pixel consists of two values that indirectly
specify the corresponding pixel coordinates in the source exemplar. This fact
enables us to use a simple look-up table to retrieve, for each target pixel,
the corresponding location in the source exemplar. In a more complex scenario
where additional guiding channels are used, we can accelerate the retrieval
using search trees~\cite{Arya98}. Once we know the corresponding source pixel,
we calculate the difference between the guidance values in local
spatially-aligned regions. The target pixels having guidance difference smaller
than a user-defined threshold belong to the current chunk~(\fg{splat}c). We
copy those corresponding pixels and paste them in the target image~(\fg{splat}d).
By repeating the searching and copying steps, we eventually cover all pixels
in the target image~(\fg{splat}e and~\fg{cmp}, left).

Our approach does not explicitly enforce textural coherence. One might expect
that seams between individual chunks will be visible. Surprisingly, for
relatively large variety of exemplars, seams are either not apparent or can be
effectively suppressed using linear blending applied around the boundaries of
individual chunks. The reasons are twofold: (1)~local guidance is often smooth
and continuous and thus two neighboring chunks are usually roughly aligned;
(2)~hand-drawn exemplars are typically highly stochastic which intrigues the
human visual system and makes the structural inconsistencies less
noticable~\cite{Ashikhmin01}.

\subsection{Implementation Details}

The basic algorithm can be implemented in a brute-force manner (see
Algorithm~\ref{algA}). Though simple, it is highly inefficient due to the
redundant visiting of target pixels and the inherent sequential nature that
prohibits parallel implementation.

\setlength{\algomargin}{0em}
\SetCommentSty{emph}
\SetFuncSty{textsf}
\SetFuncArgSty{textrm}

\begin{algorithm}
  \SetAlgoCaptionLayout{centerline}
  \caption{\CS}
  \DontPrintSemicolon
  \SetAlgoVlined
  \SetKwInOut{Input}{Inputs}
  \SetKwInOut{Output}{Output}
  \SetSideCommentRight
  \SetKwComment{KwComm}{}{}
  \Input{source style exemplar $C_S$, source guides $G_S$, target guides $G_T$, threshold $t$.}
  \Output{target stylized image $C_T$.}
  \BlankLine
  \SetKw{KwFakeFor}{for}
  \SetKw{KwFakeDo}{do}
  \SetKwFunction{FStyBlit}{\CS}
  \SetKwProg{Fn}{}{:}{}
  \Fn{\FStyBlit{}}{
    \For{each pixel $\boldsymbol{p} \in C_T$}
    {
      \BlankLine
      \If{$C_{T}[\boldsymbol{p}]$ is empty}
      {
        \BlankLine
        $\boldsymbol{u}^\star$ = $argmin_{\boldsymbol{u}}\,||G_T[\boldsymbol{p}]-G_S[\boldsymbol{u}]||$\;
        \BlankLine
        \For{each pixel $\boldsymbol{q} \in C_S$}
        {
           \BlankLine
           \If{$C_T[\boldsymbol{p}+(\boldsymbol{q}-\boldsymbol{u}^\star)]$ is empty}
           {
             \BlankLine
             $e$ = $||G_T[\boldsymbol{p}+(\boldsymbol{q}-\boldsymbol{u}^\star)]-G_S[\boldsymbol{q}]||$\;
             \BlankLine
             \If{$e<t$}
             {
              $C_T[\boldsymbol{p}+(\boldsymbol{q}-\boldsymbol{u}^\star)] = C_S[\boldsymbol{q}]$\;
             }
           }
        }
      }
    }
  }
\label{algA}
\end{algorithm}

\begin{figure}[h]
\def\svgwidth{\hsize}\import{}{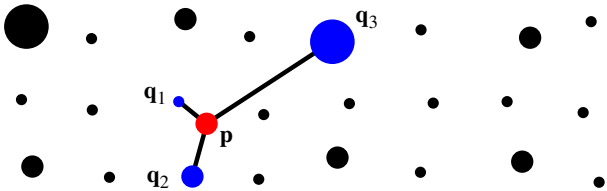}\caption{An example hierarchy of spatially distributed seeds
$\boldsymbol{q}_l$ (black and blue dots). The hierarchy level $l$ corresponds
to size of the dots: the dots in the top level are the largest. For every
target pixel $\boldsymbol{p}$ (red dot), we proceed from the top level to the
bottom $l=\{3, 2, 1\}$. At the top level, we retrieve the spatially nearest
seed ${q}_3$, and check whether the guidance value between $p$ and ${q}_3$
falls bellow a specified threshold. If not, we proceed to the nearest seed in
the next lower level ${q}_2$ and then ${q}_1$. }\label{fig:hierarchy}
\end{figure}

\begin{algorithm}
  \SetAlgoCaptionLayout{centerline}
  \caption{Parallel\CS}
  \DontPrintSemicolon
  \SetAlgoVlined
  \SetKwInOut{Input}{Inputs}
  \SetKwInOut{Output}{Output}
  \SetSideCommentRight
  \SetKwComment{KwComm}{}{}
  \Input{target pixel $\boldsymbol{p}$, target guides $G_T$, source guides $G_S$, source style exemplar $C_S$, threshold $t$, number of levels $L$.}
  \Output{stylized target pixel color $C_T[\boldsymbol{p}].$}
  \BlankLine
  \SetKw{KwBreak}{break}
  \SetKw{KwFakeFor}{for}
  \SetKw{KwFakeDo}{do}
  \SetKwFunction{FSeedPoint}{SeedPoint}
  \SetKwProg{Fn}{}{:}{}
  \Fn{\FSeedPoint{pixel $\boldsymbol{p}$,seed spacing $h$}}
  {
    $\boldsymbol{b}$ = $\left \lfloor \boldsymbol{p}/h \right \rfloor$;\,\,\,$\boldsymbol{j}$ = \texttt{RandomJitterTable}[$\boldsymbol{b}$]\;
    \KwRet $\left \lfloor h \cdot (\boldsymbol{b}+\boldsymbol{j}) \right \rfloor$\;
  }
  \BlankLine
  \SetKwFunction{FNearestSeed}{NearestSeed}
  \SetKwProg{Fn}{}{:}{}
  \Fn{\FNearestSeed{pixel $\boldsymbol{p}$,seed spacing $h$}}{
    $d^\star$ = $\infty$\;
    \KwFakeFor $x \in \{-1,0,+1\}$ \KwFakeDo\;
    \For{$y \in \{-1,0,+1\}$}
    {
      $\boldsymbol{s}$ = \FSeedPoint{$\boldsymbol{p}+h\cdot(x,y)$,$h$}\;
      $d$ = $||\boldsymbol{s}-\boldsymbol{p}||$\;
      \If{$d<d^\star$}
      {
        $\boldsymbol{s}^\star$ = $\boldsymbol{s}$;\,\,\,$d^\star$ = $d$\;
      }
    }
    \KwRet $\boldsymbol{s}^\star$\;
  }
  \BlankLine
  \SetKwFunction{FStyBlit}{Parallel\CS}
  \SetKwProg{Fn}{}{:}{}
  \Fn{\FStyBlit{pixel $\boldsymbol{p}$}}{
    \For{each level $l \in (L,\ldots,1)$}
    {
      $\boldsymbol{q}_l$ = \FNearestSeed{$\boldsymbol{p}$,$2^{l}$}\;
      $\boldsymbol{u}^\star$ = $argmin_{\boldsymbol{u}}\,||G_T[\boldsymbol{q}_l]-G_S[\boldsymbol{u}]||$\KwComm*[r]{\footnotesize{$\leftarrow$ found via lookup,}}
      $e$ = $||G_T[\boldsymbol{p}]-G_S[\boldsymbol{u}^\star+(\boldsymbol{p}-\boldsymbol{q}_l)]||$\KwComm*[r]{\footnotesize{or a tree search.}}
      \If{$e<t$}
      {
        $C_T[\boldsymbol{p}] = C_S[\boldsymbol{u}^\star+(\boldsymbol{p}-\boldsymbol{q}_l)]$\;
        \KwBreak\;
      }
    }
  }
\label{algB}
\end{algorithm}

To overcome the mentioned drawbacks, we use a more efficient approach that is
fully parallel and guarantees that every target pixel will be visited only once
(see Algorithm~\ref{algB}). The key idea here is to define an implicit
hierarchy of target seeds $\boldsymbol{q}$ (see~\fg{hierarchy}) with a
different granularity. On the top level, seeds are distributed randomly far
apart. On the lower levels, the distance between them is gradually decreased by
a factor of 2. Algorithmically we build this hierarchy by placing dots at
regular grid points whose positions are randomly perturbed. Then for every
target pixel $\boldsymbol{p}$, we start at the top level of our seed hierarchy
and find the spatially nearest target seed $\boldsymbol{q}_l$ within the same
level $l$.

If the nearest seed yields guidance error below a specific threshold, we
transfer the corresponding style color to the target pixel and stop the
traversal, otherwise we enter the next lower level of the hierarchy and
continue until we reach the bottom level.

When seams become obvious, we can optionally perform blending on the boundaries
of individual chunks. This can be simply implemented by replacing the transfer
of pixel colors with the transfer of pixel coordinates, i.e., every target
pixel will be assigned its corresponding source pixel coordinates. This
structure is equivalent to the nearest neighbour field used in patch-based
synthesis. Then, the final colors are obtained using a voting step~\cite{Kwatra05,Wexler07}
where the color of every target pixel is computed as the average color of
co-located pixels from a set of source patches that intersect the currently
processed target pixel. This operation is simple to implement and is in fact
equivalent to performing blending only at chunk boundaries.

\subsection{Extension to Animation}

Our approach can also be extended to animations. The local guidance implicitly
encourages temporal coherence in the synthesized content while the
randomization of seed points slightly perturbs the structure of the resulting
mosaic. This creates a slight temporal flickering effect which gives the
observer an illusion of a hand-colored animation where every frame is drawn
independently by hand~\cite{Fiser14}. Moreover, the amount of flickering can be
controlled by changing the guidance threshold. Higher threshold gives rise to
larger chunks and more visible visual changes between consecutive frames and
thus the amount of flickering is increased.

\section{Results}
\label{sec:results}

We implemented our approach on the CPU using C++ and on the GPU using OpenGL
with GLSL (for desktop) as well as WebGL (for mobile devices). On a single core
CPU (Core i7, 2.8 GHz) we stylize a one-megapixel image at 10 frames per second
while on the GPU (GeForce GTX 970) we can achieve more than 100 frames per
second at 4K resolution. This represents three orders of magnitude speedup as
compared to the original StyLit algorithm~\cite{Fiser16} which requires
computationally demanding iterative optimization. Such improvement enables us
to perform real-time style transfer even on devices with lower computational
budget including mid-range mobile phones (using WebGL 1.0 we can achieve, e.g.,
15 frames per second full screen on the Samsung Galaxy A3).

\begin{figure*}[ht!]
\def\svgwidth{\hsize}\import{}{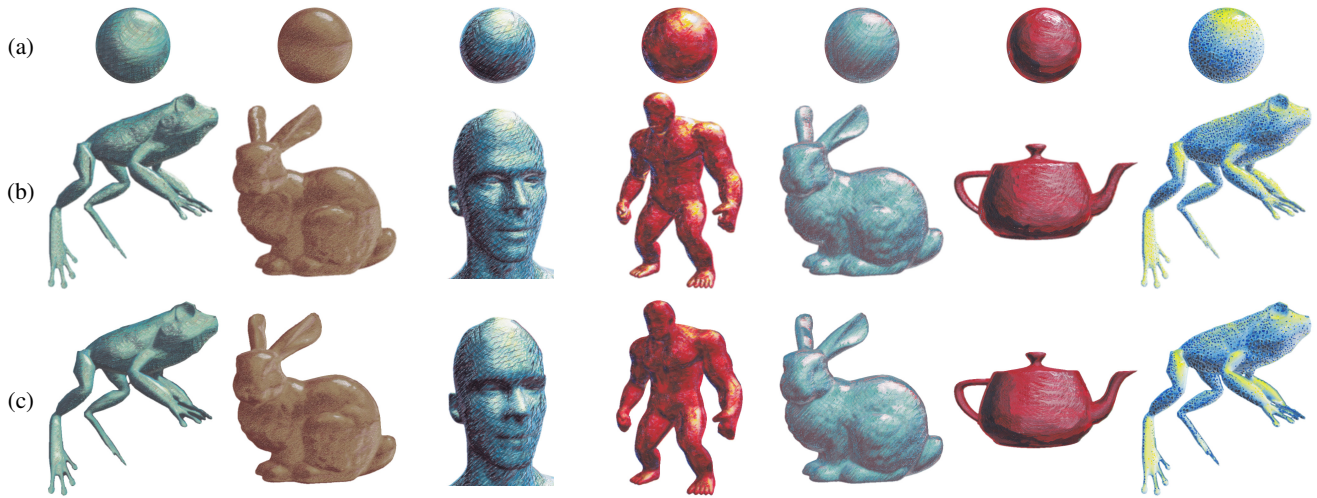}\caption{Comparison with StyLit~\protect\shortcite{Fiser16}: original
style exemplar~(a), the result of our approach~(b), and the result of StyLit~(c).}\label{fig:res}
\end{figure*}

\begin{figure}[h]
\def\svgwidth{\hsize}\import{}{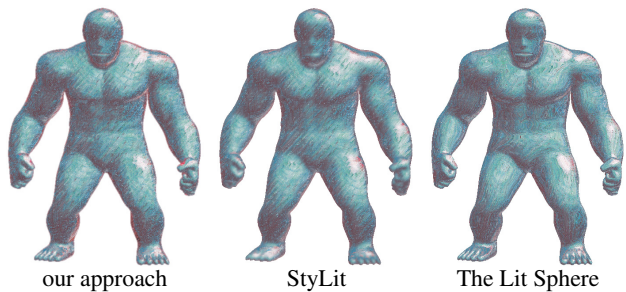}\caption{Stylized results produced by our method~(left),
StyLit~\cite{Fiser16}~(middle) and The Lit Sphere~\cite{Sloan01}~(right).
Compared to StyLit, our approach is orders of magnitude faster and produces
similar result quality without explicitly enforcing textural coherence.
Compared to The Lit Sphere, our algorithm is equally fast, but better preserves
the high-level structure of the used artistic media; i.e., large directional
brush strokes are better preserved.}\label{fig:cmp}
\end{figure}

\begin{figure*}[ht!]
\def\svgwidth{\hsize}\import{}{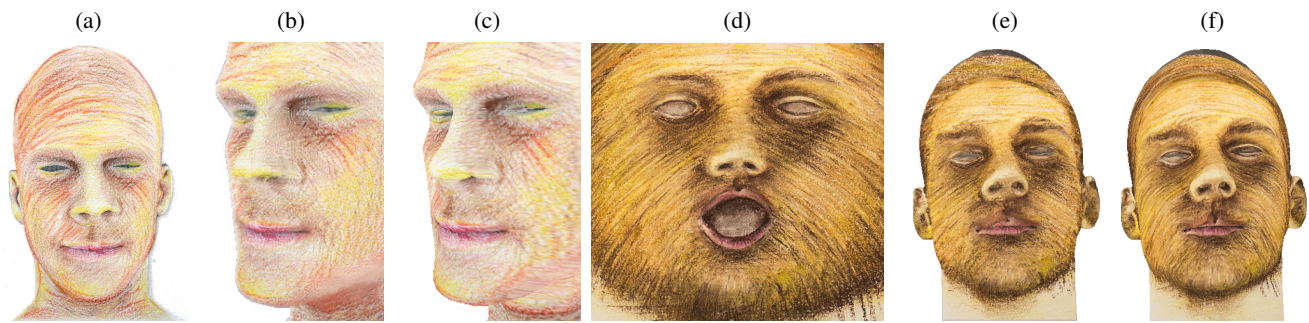}\caption{Comparison with texture mapping: original
artwork~(a, d), new viewpoint generated using our approach~(b, e) and
using texture mapping~(c, f).}\label{fig:tex}
\end{figure*}

We tested our approach in three different style transfer scenarios where local
guidance is used: normals (see~\fg{res}), texture coordinates (see~\fg{cmp}
and~\ref{fig:tex}), and a displacement field (see~\fg{fs}). For additional
results see also~\fg{teaser} and the supplementary material.

\begin{figure}[h]
\def\svgwidth{\hsize}\import{}{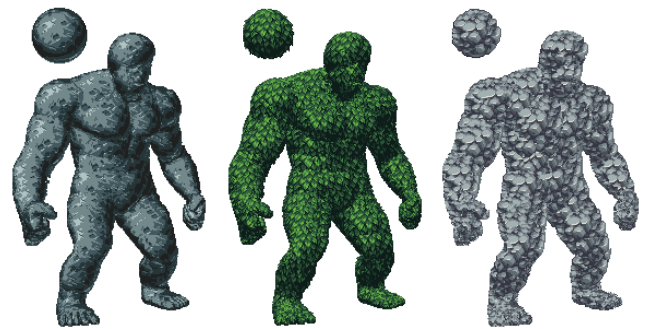}\caption{Examples of stylization where normal-based guidance is used
to transfer delicate pixel art styles. In this scenario, copy-and-paste nature
of our approach is crucial as it allows to retain essential details on the
pixel level which are important to preserve the fidelity of images that has
been created manually pixel by pixel.}\label{fig:pix}
\end{figure}

For normal-based guidance, we compared our approach with the StyLit
algorithm~\cite{Fiser16} to confirm that we produce comparable results that
preserve visually important characteristics of artistic media (see~\fg{teaser},
\ref{fig:cmp}, \ref{fig:res}, and the supplementary material that includes
results of a perceptual study). In addition, our approach also better preserves
geometric details (cf., e.g., head result in~\fg{res}) since it compares
guidance channels per pixel and does not involve any patch-based averaging used
in the StyLit algorithm. Such averaging acts as a low-pass filter applied on
the guidance channel. In the supplementary video, we present a recording of an
interactive session (on the GPU as well as on a smart phone) where the user
manipulates and animates a 3D model on which a selected artistic style is
transfered in real-time. We also demonstrate controllable temporal flickering
effect following the concept of Fi\v{s}er et al.~\shortcite{Fiser14}. Our
approach is suitable also for transferring delicate pixel art styles where even
small blurring artifacts may become apparent (see~\fg{pix}).

In the case of using texture coordinates as guidance, we compare our technique
with The Lit Sphere algorithm~\cite{Sloan01} which is based on texture mapping,
i.e., it directly maps colors between corresponding pixels according to a
one-to-one mapping specified by the normal values. Due to pixel-level
processing, distinct high-level structural features of the artistic style
become corrupted in contrast to our approach that uses larger bitmap chunks
(see~\fg{cmp} and the supplementary material). This improvement is also visible
in the case where texture coordinates are derived directly from a planar
parametrization (unwrap) of the target 3D mesh (see~\fg{teaser}, \ref{fig:tex},
and the supplementary material). Here the style exemplar can be painted on a
specific 2D projection of the 3D mesh~\cite{Magnenat15} or directly on the
planar unwrap. In both cases our approach transfers larger chunks of the
original texture which effectively removes artifacts caused by texture mapping
and better preserves the fidelity of the artistic media. To do that, however, a
larger threshold is required which can break the structure of high-level
geometric features. To avoid this artifact we include additional segmentation
guide which preventes chunks to cross boundaries of semantically important
regions (see supplementary material for examples of used guiding channels).

\begin{figure}[h!]
\def\svgwidth{\hsize}\import{}{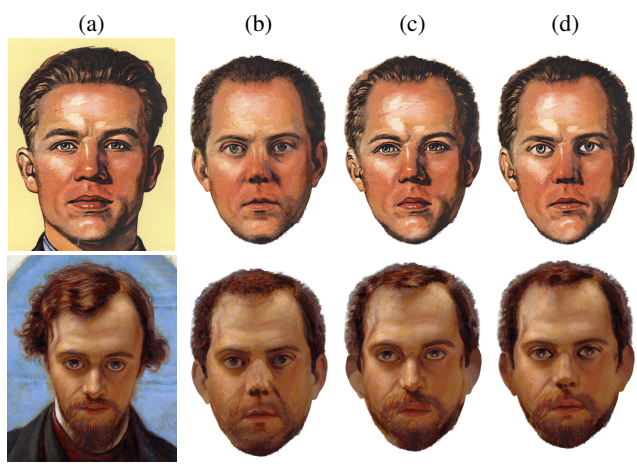}\caption{Comparison with
FaceStyle~\protect\shortcite{Fiser17}: original style exemplar~(a), the result
of our method using strong~(b) and weak~(c) appearance guide, and the result of
FaceStyle~(d).}\label{fig:fs}
\end{figure}

\begin{figure*} \includegraphics[width=\hsize]{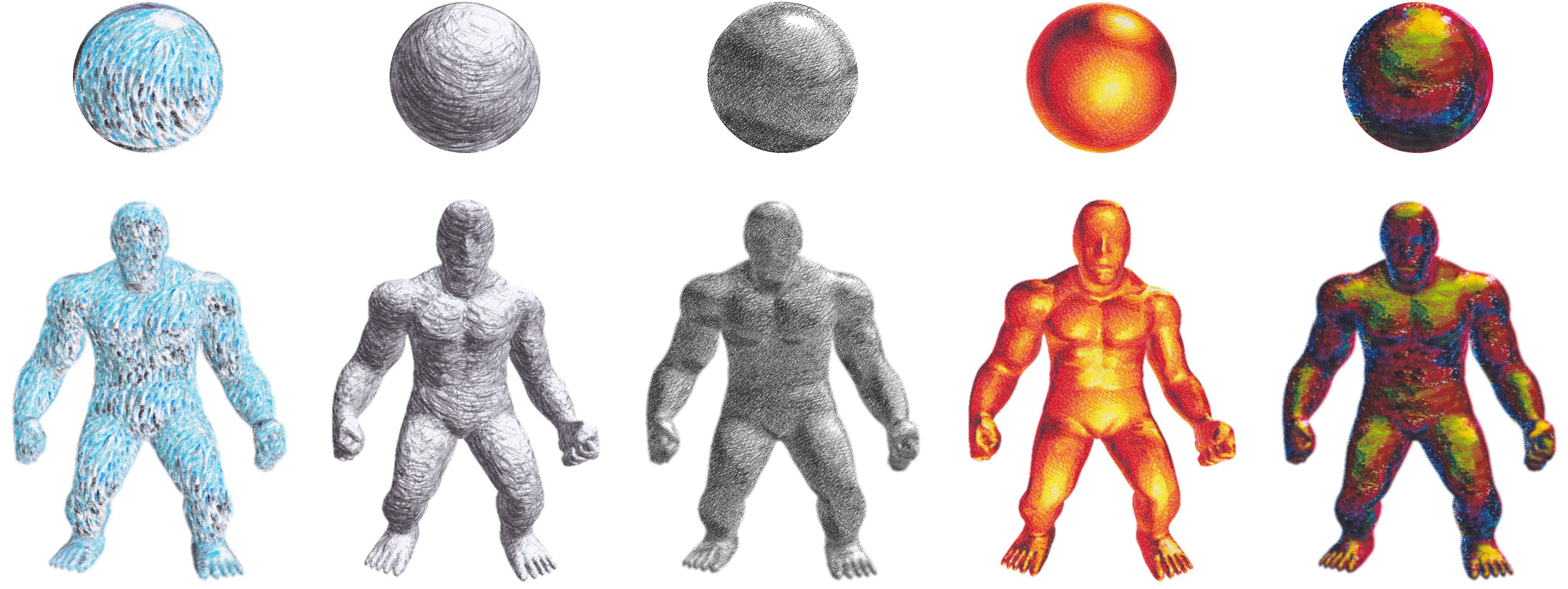}
\caption{Additional results with normal-based guidance demonstrating diversity
of style exemplars that can be used in our method: style exemplars~(top row),
result of our method~(bottom row).} \label{fig:resO}
\end{figure*}

Finally, we tested our approach in a scenario where a dense displacement field
is used as a local guide. An example of such setting is artistic style transfer
to human portraits~\cite{Fiser17}. Here the displacement field is defined by a
set of corresponding facial landmarks detected in the source exemplar and in
the target subject. Moving least squares deformation~\cite{Schaefer06} is used
to compute dense correspondences, i.e., the resulting displacement field.
Besides the local guide, two additional guidance channels are used for
patch-based synthesis: a segmentation map containing semantically important
facial parts (head, hair, eyes, eyebrows, nose, and mouth) and an appearance
guide that helps to preserve subject's identity (see the supplementary material
for examples of all guiding channels). The resulting visual quality is
comparable or a bit inferior to the previous work, but sufficient for
applications with limited computational resources (see~\fg{teaser},
\ref{fig:fs}, and the supplementary material). To demonstrate such an
application a recording of live session with real-time facial style transfer to
a streamed video is presented in the supplementary material. To highlight the
benefit of our method, the result of our algorithm is compared side-by-side
with a simple texture mapping scheme. Note, how our approach better preserves
the fidelity of the original artistic media.

\section{Limitations and Future Work}
\label{sec:limits}
Although our method produces visually pleasing results for a variety of
different style exemplars (see~\fg{resO}) and different types of guidance
there are some limitations that needs to be taken into account.

\begin{figure}[h]
\def\svgwidth{\hsize}\import{}{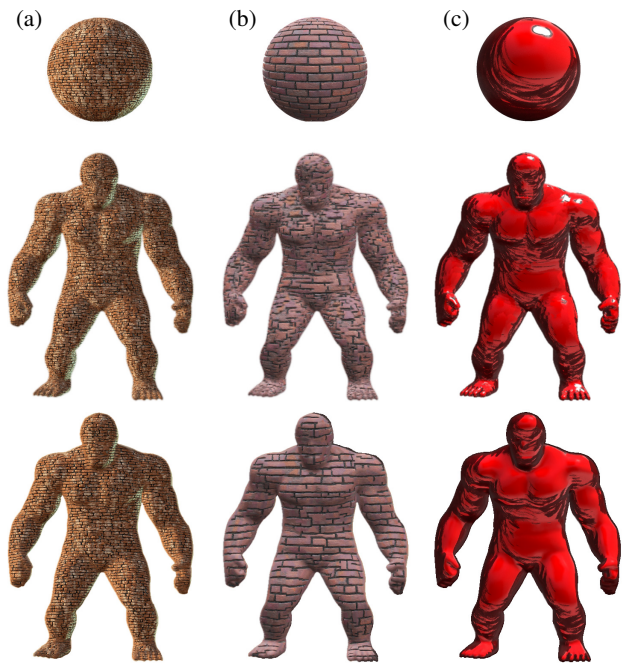}\caption{Pushing the limits: when a (semi-)regular texture is
used as a style exemplar our method may introduce visible misalignment of
regular features; when the scale of those features is comparable to the size of
individual chunks our method produces sufficient results~(a); when it becomes
larger the misalignment is more apparent~(b); when the style exemplar contains
smooth gradients as well as high-level features~(c), it is usually difficult to
find a threshold that results in both smoothness and structure preservation; in
those cases fully fledged patch-based synthesis (bottom row) usually produces
more convincing results thanks to the mechanism that encourages texture
coherence.}\label{fig:structure}
\end{figure}

Our method is suitable especially for hand-drawn style exemplars which usually
contain highly stochastic textures that helps to suppress slight misalignment
of individual chunks. For non-stochastic (semi-)regular textures like brick
wall our approach may produce visible misalignment depending on the scale of
repetitive features. Small scale bricks would work (see~\fg{structure}a) while
larger may cause artifacts (\fg{structure}b). When the style exemplar contains
smooth color gradients as well as larger coherent structures (e.g.,~distinct
brush strokes together) it can be hard to find a proper threshold which
preserve both smoothness of gradients as well as the structure of strokes. In
this scenario patch-based synthesis with textural coherence~\cite{Fiser16}
would produce better results (see~\fg{structure}d--f).

\begin{figure}[h]
\def\svgwidth{\hsize}\import{}{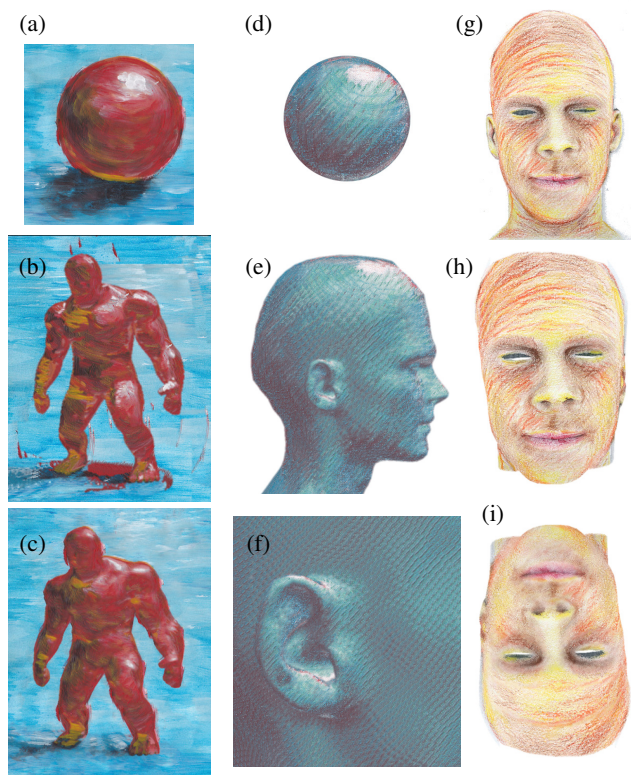}\caption{Limitations: when a set of guiding channels does
not contain local guide, for instance when light path
expressions~\cite{Fiser16} are used, our approach may introduce visible
seams~(b) in contrast to patch-based synthesis~(c); when the target contains
large areas of pixels having constant guidance values, our approach produces
visible repetition of textures~(e) which can be pronounced more when
zooming-in~(f); when the orientation of local guide changes considerably
(vertically flipped), translation cannot accommodate this change and our
technique starts to produce smaller chunks~(i), i.e., fails to preserve
high-frequency details of the style exemplar~(g) in contrast to a style
transfer with compatible orientation~(h).}\label{fig:limits}
\end{figure}

Our technique can introduce visible misalignment of individual chunks in cases
when a set of guidance channels used for the style transfer does not contain
local guide or when the influence of local guide is low as compared to other
channels. Example of such scenario can be the usage of light path expressions
in~\cite{Fiser16} (see~\fg{limits}a--c). In this case we envision as a future
work a mechanism that allows to improve the alignment of individual chunks in
the post-process.

Our approach also shares some limitations with techniques that uses full
patch-based synthesis. It may produce excessive repetition artifacts in case
the scale of the target object is fairly different as compared the object in
the style exemplar. This can happen, e.g., during zoom-in operations or when
there is not enough variability in the guidance, e.g., when stylizing flat
surface using a spherical exemplar (see~\fg{limits}d--f). As a future work we
consider to adopt a mechanism which will avoid excessive overuse of particular
pixels (e.g.~\cite{Kaspar15}).

Another limitation is related to rotation in the image plane when texture
coordinates or displacement fields are used for guidance. In this situation
corresponding counterparts of target seeds can be found easily, however, as
their neighbourhoods have notably different content caused by rotation, the
error threshold limits the size of target chunks and the method will introduce
blur into the result (see~\fg{limits}g--i). To alleviate this issue one can
pre-rotate the source guidance to match with the dominant orientation in the
target channel.

\section{Conclusion}
\label{sec:conclude}
We have presented a new approach to example-based style transfer suitable for
applications where a strong local guidance is used. We demonstrated that in
this scenario computationally demanding patch-based synthesis converge to a
solution that can be easily mimicked using relatively simple algorithm with
notably lower computation overhead. We also showed that considering textural
coherence is not crucial for successful style transfer as local guidance in
conjunction with the visual masking effectively suppresses visible seams for a
variety of hand-crafted exemplars. Since our method is several orders of
magnitude faster as compared to current state-of-the-art it enables real-time
style transfer even in applications with limited computational resources
available.

\bibliographystyle{eg-alpha-doi}

\bibliography{main}

\newcommand{\etalchar}[1]{$^{#1}$}
\def\Cadik{\v{C}ad\'{i}k}\def\Ogmen{\"{O}\u{g}men}
\begin{thebibliography}{\uppercase{SMGG01}}

\bibitem[AMN{\etalchar{*}}98]{Arya98}
\textsc{Arya S., Mount D.~M., Netanyahu N.~S., Silverman R., Wu A.~Y.}:
\newblock An optimal algorithm for approximate nearest neighbor searching in
  fixed dimensions.
\newblock \emph{Journal of the {ACM} 45}, 6 (1998), 891--923.

\bibitem[Ash01]{Ashikhmin01}
\textsc{Ashikhmin M.}:
\newblock Synthesizing natural textures.
\newblock In \emph{Proceedings of Symposium on Interactive 3D Graphics} (2001),
  pp.~217--226.

\bibitem[BCK{\etalchar{*}}13]{Benard13}
\textsc{B{\'e}nard P., Cole F., Kass M., Mordatch I., Hegarty J., Senn M.~S.,
  Fleischer K., Pesare D., Breeden K.}:
\newblock Stylizing animation by example.
\newblock \emph{ACM Transactions on Graphics 32}, 4 (2013), 119.

\bibitem[BKR17]{Bi17}
\textsc{Bi S., Kalantari N.~K., Ramamoorthi R.}:
\newblock Patch-based optimization for image-based texture mapping.
\newblock \emph{ACM Transactions on Graphics 36}, 4 (2017), 106.

\bibitem[BN76]{Blinn76}
\textsc{Blinn J.~F., Newell M.~E.}:
\newblock Texture and reflection in computer generated images.
\newblock \emph{Communications of the ACM 19}, 10 (1976), 542--547.

\bibitem[BSM{\etalchar{*}}07]{Breslav07}
\textsc{Breslav S., Szerszen K., Markosian L., Barla P., Thollot J.}:
\newblock Dynamic 2{D} patterns for shading 3{D} scenes.
\newblock \emph{ACM Transactions on Graphics 26}, 3 (2007), 20.

\bibitem[BTM06]{Barla06}
\textsc{Barla P., Thollot J., Markosian L.}:
\newblock X-toon: An extended toon shader.
\newblock In \emph{Proceedings of International Symposium on Non-Photorealistic
  Animation and Rendering} (2006), pp.~127--132.

\bibitem[BZ17]{Barnes17}
\textsc{Barnes C., Zhang F.-L.}:
\newblock A survey of the state-of-the-art in patch-based synthesis.
\newblock \emph{Computational Visual Media 3}, 1 (2017), 3--20.

\bibitem[DBP{\etalchar{*}}15]{Diamanti15}
\textsc{Diamanti O., Barnes C., Paris S., Shechtman E., Sorkine-Hornung O.}:
\newblock Synthesis of complex image appearance from limited exemplars.
\newblock \emph{ACM Transactions on Graphics 34}, 2 (2015), 22.

\bibitem[DTM96]{Debevec96}
\textsc{Debevec P.~E., Taylor C.~J., Malik J.}:
\newblock Modeling and rendering architecture from photographs: A hybrid
  geometry- and image-based approach.
\newblock In \emph{SIGGRAPH Conference Proceedings} (1996), pp.~11--20.

\bibitem[EF01]{Efros01}
\textsc{Efros A.~A., Freeman W.~T.}:
\newblock Image quilting for texture synthesis and transfer.
\newblock In \emph{SIGGRAPH Conference Proceedings} (2001), pp.~341--346.

\bibitem[EL99]{Efros99}
\textsc{Efros A.~A., Leung T.~K.}:
\newblock Texture synthesis by non-parametric sampling.
\newblock In \emph{Proceedings of IEEE International Conference on Computer
  Vision} (1999), pp.~1033--1038.

\bibitem[FJL{\etalchar{*}}16]{Fiser16}
\textsc{Fi\v{s}er J., Jamri\v{s}ka O., Luk\'{a}\v{c} M., Shechtman E., Asente
  P., Lu J., S\'{y}kora D.}:
\newblock {StyLit}: {Illumination}-guided example-based stylization of {3D}
  renderings.
\newblock \emph{ACM Transactions on Graphics 35}, 4 (2016), 92.

\bibitem[FJS{\etalchar{*}}17]{Fiser17}
\textsc{Fi\v{s}er J., Jamri\v{s}ka O., Simons D., Shechtman E., Lu J., Asente
  P., Luk\'{a}\v{c} M., S\'{y}kora D.}:
\newblock Example-based synthesis of stylized facial animations.
\newblock \emph{ACM Transactions on Graphics 36}, 4 (2017), 155.

\bibitem[FLJ{\etalchar{*}}14]{Fiser14}
\textsc{Fi\v{s}er J., Luk\'{a}\v{c} M., Jamri\v{s}ka O., \v{C}ad\'{\i}k M.,
  Gingold Y., Asente P., S\'{y}kora D.}:
\newblock {Color Me Noisy}: Example-based rendering of hand-colored animations
  with temporal noise control.
\newblock \emph{Computer Graphics Forum 33}, 4 (2014), 1--10.

\bibitem[GEB16]{Gatys16}
\textsc{Gatys L.~A., Ecker A.~S., Bethge M.}:
\newblock Image style transfer using convolutional neural networks.
\newblock In \emph{Proceedings of IEEE Conference on Computer Vision and
  Pattern Recognition} (2016), pp.~2414--2423.

\bibitem[Her98]{Hertzmann98}
\textsc{Hertzmann A.}:
\newblock Painterly rendering with curved brush strokes of multiple sizes.
\newblock In \emph{SIGGRAPH Conference Proceedings} (1998), pp.~453--460.

\bibitem[HJO{\etalchar{*}}01]{Hertzmann01}
\textsc{Hertzmann A., Jacobs C.~E., Oliver N., Curless B., Salesin D.~H.}:
\newblock Image analogies.
\newblock In \emph{SIGGRAPH Conference Proceedings} (2001), pp.~327--340.

\bibitem[JAFF16]{Johnson16}
\textsc{Johnson J., Alahi A., Fei-Fei L.}:
\newblock Perceptual losses for real-time style transfer and super-resolution.
\newblock In \emph{Proceedings of European Conference on Computer Vision}
  (2016), pp.~694--711.

\bibitem[KCWI13]{Kyprianidis13}
\textsc{Kyprianidis J.~E., Collomosse J., Wang T., Isenberg T.}:
\newblock State of the ``art'': A taxonomy of artistic stylization techniques
  for images and video.
\newblock \emph{IEEE Transactions on Visualization and Computer Graphics 19}, 5
  (2013), 866--885.

\bibitem[KEBK05]{Kwatra05}
\textsc{Kwatra V., Essa I.~A., Bobick A.~F., Kwatra N.}:
\newblock Texture optimization for example-based synthesis.
\newblock \emph{ACM Transactions on Graphics 24}, 3 (2005), 795--802.

\bibitem[KNL{\etalchar{*}}15]{Kaspar15}
\textsc{Kaspar A., Neubert B., Lischinski D., Pauly M., Kopf J.}:
\newblock Self tuning texture optimization.
\newblock \emph{Computer Graphics Forum 34}, 2 (2015), 349--360.

\bibitem[KSE{\etalchar{*}}03]{Kwatra03}
\textsc{Kwatra V., Sch{\"o}dl A., Essa I.~A., Turk G., Bobick A.~F.}:
\newblock Graphcut textures: Image and video synthesis using graph cuts.
\newblock \emph{ACM Transactions on Graphics 22}, 3 (2003), 277--286.

\bibitem[LH05]{Lefebvre05}
\textsc{Lefebvre S., Hoppe H.}:
\newblock Parallel controllable texture synthesis.
\newblock \emph{ACM Transactions on Graphics 24}, 3 (2005), 777--786.

\bibitem[LH06]{Lefebvre06}
\textsc{Lefebvre S., Hoppe H.}:
\newblock Appearance-space texture synthesis.
\newblock \emph{ACM Transactions on Graphics 25}, 3 (2006), 541--548.

\bibitem[LLX{\etalchar{*}}01]{Liang01}
\textsc{Liang L., Liu C., Xu Y.-Q., Guo B., Shum H.-Y.}:
\newblock Real-time texture synthesis by patch-based sampling.
\newblock \emph{ACM Transactions on Graphics 20}, 3 (2001), 127--150.

\bibitem[LW16]{Li16}
\textsc{Li C., Wand M.}:
\newblock Combining markov random fields and convolutional neural networks for
  image synthesis.
\newblock In \emph{Proceedings of IEEE Conference on Computer Vision and
  Pattern Recognition} (2016), pp.~2479--2486.

\bibitem[LYY{\etalchar{*}}17]{Liao17}
\textsc{Liao J., Yao Y., Yuan L., Hua G., Kang S.~B.}:
\newblock Visual attribute transfer through deep image analogy.
\newblock \emph{ACM Transactions on Graphics 36}, 4 (2017), 120.

\bibitem[MNZ{\etalchar{*}}15]{Magnenat15}
\textsc{Magnenat S., Ngo D.~T., Zünd F., Ryffel M., Noris G., Roethlin G.,
  Marra A., Nitti M., Fua P., Gross M.~H., Sumner R.~W.}:
\newblock Live texturing of augmented reality characters from colored drawings.
\newblock \emph{IEEE Transactions on Visualization and Computer Graphics 21},
  11 (2015), 1201--1210.

\bibitem[PFH00]{Praun00}
\textsc{Praun E., Finkelstein A., Hoppe H.}:
\newblock Lapped textures.
\newblock In \emph{SIGGRAPH Conference Proceedings} (2000), pp.~465--470.

\bibitem[PKVP09]{Pritch09}
\textsc{Pritch Y., Kav-Venaki E., Peleg S.}:
\newblock Shift-map image editing.
\newblock In \emph{Proceedings of IEEE International Conference on Computer
  Vision} (2009), pp.~151--158.

\bibitem[PS00]{Portilla00}
\textsc{Portilla J., Simoncelli E.~P.}:
\newblock A parametric texture model based on joint statistics of complex
  wavelet coefficients.
\newblock \emph{International Journal of Computer Vision 40}, 1 (2000), 49--70.

\bibitem[RRFT14]{Rematas14}
\textsc{Rematas K., Ritschel T., Fritz M., Tuytelaars T.}:
\newblock Image-based synthesis and re-synthesis of viewpoints guided by 3d
  models.
\newblock In \emph{Proceedings of IEEE Conference on Computer Vision and
  Pattern Recognition} (2014), pp.~3898--3905.

\bibitem[SED16]{Selim16}
\textsc{Selim A., Elgharib M., Doyle L.}:
\newblock Painting style transfer for head portraits using convolutional neural
  networks.
\newblock \emph{ACM Transactions on Graphics 35}, 4 (2016), 129.

\bibitem[SID17]{Semmo17}
\textsc{Semmo A., Isenberg T., Döllner J.}:
\newblock Neural style transfer: A paradigm shift for image-based artistic
  rendering?
\newblock In \emph{Proceedings of International Symposium on Non-Photorealistic
  Animation and Rendering} (2017), p.~5.

\bibitem[SMGG01]{Sloan01}
\textsc{Sloan P.-P.~J., Martin W., Gooch A., Gooch B.}:
\newblock {The Lit Sphere}: {A} model for capturing {NPR} shading from art.
\newblock In \emph{Proceedings of Graphics Interface} (2001), pp.~143--150.

\bibitem[SMW06]{Schaefer06}
\textsc{Schaefer S., {McPhail} T., Warren J.}:
\newblock Image deformation using moving least squares.
\newblock \emph{ACM Transactions on Graphics 25}, 3 (2006), 533--540.

\bibitem[SSGS11]{Schmid11}
\textsc{Schmid J., Senn M.~S., Gross M., Sumner R.~W.}:
\newblock Overcoat: an implicit canvas for {3D} painting.
\newblock \emph{ACM Transactions on Graphics 30}, 4 (2011), 28.

\bibitem[SWHS97]{Salisbury97}
\textsc{Salisbury M.~P., Wong M.~T., Hughes J.~F., Salesin D.~H.}:
\newblock Orientable textures for image-based pen-and-ink illustration.
\newblock In \emph{SIGGRAPH Conference Proceedings} (1997), pp.~401--406.

\bibitem[SZ14]{Simonyan14}
\textsc{Simonyan K., Zisserman A.}:
\newblock Very deep convolutional networks for large-scale image recognition.
\newblock \emph{CoRR abs/1409.1556} (2014).

\bibitem[TAY13]{Todo13}
\textsc{Todo H., Anjyo K., Yokoyama S.}:
\newblock Lit-sphere extension for artistic rendering.
\newblock \emph{The Visual Computer 29}, 6--8 (2013), 473--480.

\bibitem[WSI07]{Wexler07}
\textsc{Wexler Y., Shechtman E., Irani M.}:
\newblock Space-time completion of video.
\newblock \emph{IEEE Transactions on Pattern Analysis and Machine Intelligence
  29}, 3 (2007), 463--476.

\bibitem[ZSL{\etalchar{*}}17]{Zhou17}
\textsc{Zhou Y., Shi H., Lischinski D., Gong M., Kopf J., Huang H.}:
\newblock Analysis and controlled synthesis of inhomogeneous textures.
\newblock \emph{Computer Graphics Forum 36}, 2 (2017), 199--212.

\end{thebibliography}

\end{document}